\documentclass[final
    ,final            
  ]
  {aipproc}
\usepackage{epsf,epsfig,amsmath,amssymb,bm}
\usepackage{graphicx}
\layoutstyle{6x9}

\def\({\left(}
\def\[{\left[}
\def\){\right)}
\def\]{\right]}
\def\ex{\hbox{e}}
\def\<{\langle}
\def\>{\rangle}

\newcommand{\vecc}[1]{\mbox{\boldmath $#1$}}

\def\e{\epsilon}

\def\ex{\hbox{e}}
\def\S{\Sigma}

\def\<{\langle}
\def\>{\rangle}

\def\a{\alpha}

\def\d{\delta}

\def\m{\mu}
\def\n{\nu}

\def\({\left(}
\def\[{\left[}
\def\){\right)}
\def\]{\right]}

\begin{document}
\hfill{RUB-TPII-10/08}
\bigskip

\title{TMD PDF's: \\ gauge invariance, RG properties and Wilson lines 
\thanks{Invited talk presented by the first author at the International 
Workshop on Diffraction in High Energy and Nuclear Physics, 
La Londe-les-Maures, France, 9-14 Sept 2008}}


\author{I.O. Cherednikov}{
  address={Bogoliubov Laboratory of Theoretical Physics, JINR \\
           RU-141980 Dubna, Russia\\
           E-mail: igor.cherednikov@jinr.ru\\}
  and
}

\author{N.G. Stefanis}{
  address={Institut f\"{u}r Theoretische Physik II,
           Ruhr-Universit\"{a}t Bochum \\
           D-44780 Bochum, Germany\\
           E-mail: stefanis@tp2.ruhr-uni-bochum.de}
}

\begin{abstract}
The UV divergences associated with transverse-momentum dependent
(TMD) parton distribution functions (PDF) are calculated together
with the ensuing one-loop anomalous dimensions in the light-cone
gauge.
Time-reversal-odd 
effects in the anomalous dimensions are
observed and the role of Glauber gluons is discussed.
A generalized renormalization procedure of TMD PDFs is proposed,
relying upon the renormalization of contour-dependent operators with
obstructions.
\end{abstract}

\maketitle


\paragraph{UV-divergences of unintegrated PDF and one-loop
anomalous dimension}

Parton distribution functions encode important information about the
inner structure of hadrons in terms of their constituents---partons
\cite{Col03,BR05,Col08}.
In inclusive processes (e.g., DIS), integrated PDFs appear which
depend on the longitudinal fraction of the momentum $x$ and the
scale of the hard subprocess $Q^2$.
Their renormalization properties are governed by the DGLAP equation.
On the other hand, the study of semi-inclusive processes, such as
SIDIS, or the Drell-Yan process---where the transverse momentum of
the produced hadrons can be observed---requires the introduction of
more complicated quantities, i.e., unintegrated, or TMD, PDFs.
Their gauge-invariant definition reads ($\xi^{+}=0$)
\cite{JY02,BJY03,BMP03}
$$
   f_{q/q}(x, \mbox{\boldmath$k_\perp$})
 =
  \frac{1}{2}
   \int \frac{d\xi^- d^2
   \vecc \xi_\perp}{2\pi (2\pi)^2}
   {\rm e}^{- i k^+ \xi^- +i \mbox{\footnotesize \boldmath$k_\perp$}
   \cdot \mbox{\footnotesize \boldmath$\xi_\perp$}}
   \left\langle  q(p) |\bar \psi (\xi^-, \xi_\perp)
   [\xi^-, \mbox{\boldmath$\xi_\perp$};
   \infty^-, \mbox{\boldmath$\xi_\perp$}]^\dagger \right.
$$
\begin{equation}
\left.
\times
   [\infty^-, \mbox{\boldmath$\xi_\perp$};
   \infty^-, \mbox{\boldmath$\infty_\perp$}]^\dagger \gamma^+
[\infty^-, \mbox{\boldmath$\infty_\perp$};
   \infty^-, \mbox{\boldmath$0_\perp$}]
   [\infty^-, \mbox{\boldmath$0_\perp$}; 0^-,\mbox{\boldmath$0_\perp$}]
   \psi (0^-,\mbox{\boldmath$0_\perp$}) |q(p)\right\rangle \
   \ ,
\label{eq:tmd_definition}
\end{equation}
where gauge invariance is restored via Wilson lines (gauge links)
with the generic form
$
{ [y,x|\Gamma] }
=
  {\cal P} \exp
  \left[-ig\int_{x[\Gamma]}^{y}dz_{\mu} A_{a}^{\mu}(z) t_{a}
  \right].
$
Note that the transverse gauge links at light-cone infinity in
(\ref{eq:tmd_definition}) contribute only in the light-cone gauge and
coerce the cancelation of the pole-prescription dependence.

However, in Eq.\ (\ref{eq:tmd_definition}), UV divergences arise,
which are associated with the features of the light-cone
gauge (or the lightlike Wilson lines) that should be cured
(see, e.g., \cite{Col08,CRS07,Bacch08,CS07}).
These divergences can be avoided by using non-lightlike gauge links in
covariant gauges, or in an axial gauge off the light cone
\cite{CS81,JMY04}.
This involves the introduction of an extra rapidity parameter and
entails an additional evolution equation \cite{CS81}, rendering the
reduction to the integrated PDF questionable.
In this presentation, we describe another strategy based on a
subtraction formalism of these extra divergences by means of a ``soft''
factor, defined as the vacuum average of particular Wilson lines
(and demonstrated explicitly in a covariant gauge at the one-loop level
\cite{CH00,Hau07}--see also \cite{CM04}.

Using this stratagem, we calculate the UV singularities of the
function given by Eq.\ (\ref{eq:tmd_definition}) in the light-cone
gauge.
The one-gluon exchanges, contributing to the UV-divergences, are
represented by the diagrams $(a)$ and $(b)$ in Fig.\ 1.
The source of the uncertainties and extra divergences is the pole
structure of the gluon propagator in the light-cone (LC) gauge:
$
  D^{\m\n}_{\rm LC} (q)
=
  \frac{-i}{q^2} \[
  g^{\m\n} - { \frac{q^\m n^{-\n}}{[q^+]}
              - \frac{q^\n n^{-\m} }{[q^+]} } \] \ ,
$
where $[q^+]$ bears the pole prescription in the denominator.
We consider the following pole prescriptions
\begin{equation}
      \frac{1}{[q^+]}_{\rm PV}
      =
      \frac{1}{2} \( \frac{1}{q^+ + i \eta} + \frac{1}{q^+ - i \eta} \) \ \
      \hbox{and} \ \
  \frac{1}{[q^+]}_{\rm Adv/Ret}
  =
  \frac{1}{q^+ \mp i \eta} \ ,
\label{eq:pole}
\end{equation}
keeping $\eta$ small, but finite, and using dimensional regularization to
control UV singularities.
The UV divergent part of diagrams $(a)$ and $(b)$ (without ``mirror''
contributions) is
\begin{equation}
  {\S}^{UV}_{\rm left} (p, \a_s ; \e)
=
{
  - \frac{\a_s}{\pi} \ C_{\rm F} \  \frac{1}{\e}
  \[- \frac{3}{4} -  \ln \frac{\eta}{p^+} + \frac{i\pi}{2}
  + i  \pi \ C_\infty \]
  + \a_s \ C_{\rm F} \ \frac{1}{\e} \[i C_\infty\] } \ ,
\label{eq:left}
\end{equation}
where the numerical factor $C_{\infty}$ accumulates the
pole-prescription uncertainty, and is defined as
$
 C_\infty^{\rm Adv} = 0 \ , \ C_\infty^{\rm Ret} = -1 \ ,
 C_\infty^{\rm PV} = -1/2
$.
One immediately observes that the prescription dependence is canceled
due to the contribution of the transverse gauge link at the
light-cone infinity---diagram ($b$).
Taking into account the conjugate ``mirror'' contributions, one gets
the total real UV divergent part:
\begin{equation}
  \S_{\rm tot}^{UV} (p, \a_s (\m) ; \e )
=
  \S_{\rm left} + \S_{\rm right}
=
  { - \frac{ \a_s}{4\pi} \ C_{\rm F} \ \frac{2}{\e}
  \(- 3 - 4 \ln \frac{\eta}{p^+} \) } \ .
\label{eq:tot_uv}
\end{equation}

\begin{figure}
  \includegraphics[height=.395\textheight,angle=90]{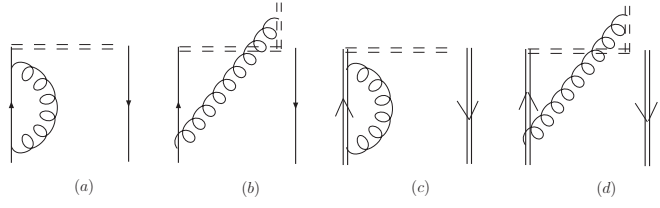}
  \caption{One-gluon exchanges ìn the TMD PDF in the light-cone gauge.
  Diagrams $(a)$ and $(b)$ give rise to UV divergences, while
  $(c)$ and $(d)$ correspond to the soft factor,
  cf.\ (\ref{eq:soft_factor_1}).
  Double lines denote gauge links and curly lines gluon propagators.
  The Hermitian conjugate, i.e., ``mirror'', diagrams are omitted. }
\end{figure}

The associated one-loop anomalous dimension is given by
\begin{equation}
  \gamma_{\rm LC}
=
  \frac{\a_s}{\pi}C_{\rm F} \(\frac{3}{4} + \ln \frac{\eta}{p^+} \)
=
  \gamma_{\rm smooth} - { \d \gamma } \ \ , \ \ \gamma_{\rm smooth}
=
  { \frac{3}{4} \frac{\a_s}{\pi}C_{\rm F} } + O(\a_s^2) \ .
\end{equation}
The defect of the anomalous dimension $\delta \gamma$ marks the
deviation of the calculated quantity from the anomalous dimension
of the gauge-invariant two-quark correlator with a gauge link along a
straight line (i.e., the connector \cite{Ste83}).
Note that $\gamma_{\rm LC}$ contains an undesirable $p^+$-dependent
term that should be removed.
The key observation here is that
$p^+ = (p \cdot n^-) \sim \cosh \chi$
defines, in fact, an angle $\chi$ between the direction of the
quark momentum $p_\mu$ and the lightlike vector $n^-$.
In the large $\chi$ limit,
$\ln p^+ \to \chi , \ \chi \to \infty$.
Thus, the defect of the anomalous dimension, $\delta \gamma$,
can be identified at the one-loop level with the cusp anomalous
dimension \cite{KR87}.
The validity of this finding at higher orders is yet unknown.

\paragraph{$T$-odd effects in the anomalous dimension}

{Time-reversal-odd} effects arise when the dependence on the intrinsic
transverse motion of partons is taken into account.
This happens in semi-inclusive processes, like SIDIS (or DY) and are
responsible for single-spin asymmetries (see, e.g., \cite{Mul08}).
The origin of such effects are the Wilson lines in the operator
definition of the TMD PDFs, since they accumulate information about
initial/final state interactions.
Our analysis shows that $T-$odd phenomena reveal themselves
also in the anomalous dimensions.
In fact, the imaginary term,
$
 {\rm Im} \ {\S}^{UV}_{\rm left}
=
 - \frac{\a_s}{2\e}C_{\rm F}
 $,
in Eq.\ (\ref{eq:left}) stems from the infinitesimal deformation
of the integration contour to circumvent the pole in the gluon
propagator subject to the pole prescriptions in (\ref{eq:pole}).
It corresponds to the imaginary term one would obtain with lightlike
Wilson lines in a covariant gauge.
In this latter case, the leading term in the Wilson line produces,
after Fourier transforming it, a similar $q^+$-pole in the denominator:
\begin{equation}
  \int_0^\infty\! d\xi^- A^+ (\xi^-, 0^+, \vecc 0_\perp)
=
  \int\! d^4 q \tilde A^+ (q)
  \int_0^\infty\! d\xi^- \ex^{-i (q^+ - i\eta)\xi^-}
=
  \int\! d^4 q \tilde A^+ (q) \frac{-i}{q^+ - i\eta} \ .
   \label{eq:cov}
\end{equation}
Taking into account that $T-$reversal corresponds to the inversion
of the Wilson line's direction flipping the sign in the denominator
from $\eta \to - \eta$, one may conclude that the origin of the $T-$odd
effects in the TMD PDFs can be traced back to their {\it local}
RG properties expressed via their anomalous dimensions.
Note that the imaginary terms in the anomalous dimensions can be
attributed to the contributions of gluons in the Glauber regime, where
their momenta are mostly transverse \cite{KR87}.
Indeed, it was recently shown
(in the Soft Collinear Effective Theory) that exactly the Glauber
gluons contribute most to the transverse gauge link,
underlying $T-$odd effects in the light-cone gauge \cite{IdMa08}.

\paragraph{Generalized definition of TMD PDFs}

In order to cancel the anomalous dimension defect $\delta \gamma$,
we introduce the counter term \cite{CH00}
\begin{equation}
  R
\equiv
 \Phi (p^+, n^- | 0) \Phi^\dagger (p^+, n^- | \xi) \ ,
\label{eq:soft_factor_1}
\end{equation}
where
$
  \Phi (p^+, n^- | \xi )
 =
  \left\langle 0
  \left| {\cal P} \exp\Big[ig \int_{\Gamma_{\rm cusp}}d\zeta^\mu
  \ t^a A^a_\mu (\xi + \zeta)\Big]
  \right|0
  \right\rangle
$
and evaluate it along
\begin{equation}
\Gamma_{\rm cusp} : \ \ \zeta_\mu
=
  \{ [p_\mu^{+}s \ , \ - \infty < s < 0] \
 \cup \ [n_\mu^-  s' \ ,
  \ 0 < s' < \infty] \ \cup \
  [ \mbox{\boldmath$l_\perp$} \tau , \, \ 0 < \tau < \infty ] \}
\label{eq:gpm}
\end{equation}
with $n_\mu^-$ being the minus light-cone vector.
Employing the renormalization techniques for contour-dependent
operators with obstructions (cusps, or self-intersections), that
induce an angle dependence \cite{Pol80,CD80,KR87}, we compute
the extra renormalization constant associated with this soft counter
term and show that it cancels the anomalous-dimension defect,
$\delta \gamma$, \cite{CS07}.
The one-loop gluon virtual corrections, contributing to the UV
divergences of $R$, are shown in Fig.\ 1 diagrams $(c)$, $(d)$.
For the UV divergent term we obtain
\begin{equation}
  \Sigma_{R}^{\rm UV}
=
  - \frac{ \alpha_s}{\pi} C_{\rm F} \   \frac{2}{\epsilon} \
  \ln \frac{\eta}{p^+}
\end{equation}
and observe that this expression equals, but with the opposite sign,
the unwanted term in front of the UV singularity related to the cusped
contour, calculated above.

Therefore, we propose to redefine the conventional TMD PDF and
absorb the soft counter term in its definition according to
\begin{equation}
f_{q/q}^{\rm mod}\left(x, \mbox{\boldmath$k_\perp$}
                 \right)
=
  \frac{1}{2}
  \int \frac{d\xi^- d^2\mbox{\boldmath$\xi_\perp$}}{2\pi (2\pi)^2}
  {\rm e}^{- i k^+ \xi^- +i \mbox{\footnotesize \boldmath$k_\perp$}
  \cdot \mbox{\footnotesize \boldmath$\xi_\perp$}}
  F(\xi^{-}, \mbox{\boldmath$\xi_\perp$})
  R (p^+, n^-|\xi^{-}, \mbox{\boldmath$\xi_\perp$}) \ ,
\label{eq:tmd_re-definition}
\end{equation}
where $F(\xi^{-}, \mbox{\boldmath$\xi_\perp$})$ denotes the matrix
element in 
(\ref{eq:tmd_definition}).
We verified \cite{CS07} that integrating over $\vecc k_\perp$
yields an integrated PDF obeying the DGLAP equation and collinear
factorization.

\paragraph{Conclusions}

The anomalous dimension of the TMD PDF in the {light-cone gauge}
was calculated in the one-loop order.
It was shown explicitly, how the transverse semi-infinite gauge link
eliminates the dependence on the pole prescriptions
in the gluon light-cone-propagator.
The T-odd effects could be linked to contributions of Glauber gluons
which induce an imaginary term in the anomalous dimension.
A generalized renormalization procedure of the TMD PDFs
was proposed, employing techniques pertaining to the renormalization
of Wilson exponentials with cusped gauge contours.


\paragraph{Acknowledgments}
This work was supported in part by the Alexander von Humboldt
Stiftung, the Deutsche Forschungsgemeinschaft (grant
436 RUS 113/881/0), the Russian Federation President's grant
1450-2003-2, and the Heisenberg--Landau Program 2008.

\end{document}